\documentclass[english,twocolumn,superscriptaddress]{revtex4-1}
\usepackage[LGR,T1]{fontenc}
\usepackage[latin9]{inputenc}
\usepackage{babel}
\usepackage{textcomp}
\usepackage{amstext}
\usepackage{graphicx}
\usepackage{subscript}
\usepackage[unicode=true,pdfusetitle,
 bookmarks=true,bookmarksnumbered=false,bookmarksopen=false,
 breaklinks=false,pdfborder={0 0 1},backref=false,colorlinks=false]
 {hyperref}

\makeatletter

\DeclareRobustCommand{\greektext}{%
  \fontencoding{LGR}\selectfont\def\encodingdefault{LGR}}
\DeclareRobustCommand{\textgreek}[1]{\leavevmode{\greektext #1}}
\DeclareFontEncoding{LGR}{}{}
\DeclareTextSymbol{\~}{LGR}{126}

\makeatother

\begin{document}

\title{Proximity effect on hydrodynamic interaction between a sphere and
a plane measured by Force Feedback Microscopy at different frequencies.}

\author{Simon Carpentier}

\affiliation{Univ. Grenoble Alpes, F-38000 Grenoble, France }

\affiliation{CNRS, Inst NEEL, F-38042 Grenoble, France}

\author{Mario S. Rodrigues}

\affiliation{Uniservity of Lisboa, Falculty of Science, BioISI-Biosystems \& Integrative
Sciences Institute, Campo Grande, Lisboa, Portugal}

\author{Elisabeth Charlaix}

\affiliation{CNRS, LIPhy, Grenoble, F-38402, France }

\author{Joël Chevrier}

\affiliation{Univ. Grenoble Alpes, F-38000 Grenoble, France }

\affiliation{CNRS, Inst NEEL, F-38042 Grenoble, France}
\begin{abstract}
In this article, we measure the viscous damping G'', and the associated
stiffness G', of a liquid flow in sphere-plane geometry in a large
frequency range. In this regime, the lubrication approximation is
expected to dominate. We first measure the static force applied to
the tip. This is made possible thanks to a force feedback method.
Adding a sub-nanometer oscillation of the tip, we obtain the dynamic
part of the interaction with solely the knowledge of the lever properties
in the experimental context using a linear transformation of the amplitude
and phase change. Using a Force Feedback Microscope (FFM)we are then
able to measure simultaneously the static force, the stiffness and
the dissipative part of the interaction in a broad frequency range
using a single AFM probe. Similar measurements have been performed
by the Surface Force Apparatuswith a probe radius hundred times bigger.
In this context the FFM can be called nano-SFA.
\end{abstract}
\maketitle
In this paper, we aim to measure the hydrodynamic interaction in water
between a microsphere and a hard and plane surface at different distances
and at different frequencies using a single probe. In this regime,
the viscous behavior is expected to be dominating. Previous experiments
have been run in air with conservative forces\cite{carpentier2014variable}.
Using an electrical coupling between an AFM tip and a metallic surface
as test interaction in air, we probed the static and dynamic part
of the interaction where in this case the stiffness is expected to
match the minus of the derivative of the static force and the damping
to be null. In liquid, from the dynamic transition from a viscous-dominated
behavior at large distance to an elastic-dominated behavior at short
distance, we aim to extract mechanical properties of soft thin films
as SFA does\cite{leroy2011hydrodynamic}. The hydrodynamic pressure
will gently deform the soft sample without touching it and propose
an alternative to the classical hard mechanical contact. The increase
in the frequency range of this mechanical testing and the decrease
of the tip radius, compare to the SFA\cite{Israelachvili1992}, allows
us to probe the viscoelastic properties of soft thin films and the
possibility of the frequency dependence and compare it to bulk values.
Properties can dramatically change with respect to the frequency as
we have demonstrated in the case of capillary bridge\cite{carpentier2015out}.
Here in this test experiment, we use non-deformable surface and spherical
probe, and we explore the dissipation due to the flowing liquid. Other
experiments with an AFM setup have been performed\cite{benmouna2002hydrodynamic,maali2005hydrodynamics}
in order to measure the hydrodymanic behavior either with respect
to the sphere-plane separation or for different frequencies using
one lever. The dissipation is expected to follow the lubrication approximation
and to be frequency dependent, f=\textgreek{w}/2\textgreek{p}, if
reported in N/m. The lubrication approximation holds for low Reynolds
number, thus R>\textcompwordmark{}>z, and the dissipation follow\cite{leroy2011hydrodynamic}:
\[
G"=\frac{6\pi\omega\eta R^{2}}{z}\;(1)
\]
 R is the tip radius, the dynamic viscous coefficient of the water,
equal to 1 mPA.s at 20\textdegree C, from IAPWS standards, and z is
the tip sample distance. The linear frequency dependency and the z\textsuperscript{-1}
behavior of the dissipation is thus a perfect test interaction in
liquid as the theory and the experiment are directly compared without
adjustable parameters. \\

The Force Feedback Microscope\cite{rodrigues2012atomic}, based on
standard AFM setup using optical fiber interferometric detection\cite{rugar1989improved},
is based on one key point: a piezoelectric element is placed on the
chip and a feedback loop keep the DC position of the tip constant
during all the measurement by canceling the static force acting on
the tip. The static feedback force is then the displacement of the
piezoelement times the lever stiffness k. The total force acting on
the tip is then equal to zero during all the measurement. This leads
to the suppression of mechanical instabilities, such as when in air
a capillary bridge nucleates, and allows us to apply subnanometers
oscillations and lets the system in the linear response regime. Thus
through linear transformation of the amplitude and phase change using
the following equations we can recover the dynamic part of the interaction
(see more details in Ref.2):

\[
G'=Fr[ncos(\Phi)-cos(\Phi\infty)]\;(2)
\]

\[
G''=Fr[\text{\textminus}nsin(\text{\textgreek{F}})+sin(\text{\textgreek{F}\ensuremath{\infty}})]\;(3)
\]

n is the normalized amplitude of oscillation, it is equal to 1 far
from the sample and \textgreek{F} is the phase. Parameters Fr and
\textgreek{F}\ensuremath{\infty} are obtained using the lever properties
in the environmental experiment and the following equations (see more
details in Ref.\cite{rodrigues2014system}):

\[
Fr=[(k-m\omega^{2})^{2}+\gamma^{2}\omega^{2}]^{1/2}
\]

\[
\text{\textgreek{F}\ensuremath{\infty}}=arctan[(\gamma/m)\omega/(\omega^{2}\text{\textendash}\omega{}_{0}^{2})]
\]
Using the phase response of the system in liquid far from the sample,
see Fig.1, we extract the resonance frequency f\textsubscript{0}
and the quality factor Q, by fitting the measured phase response by
the theoretical phase of a harmonic oscillator. The mass of the system
m is deduced from the knowledge of the resonance frequency and the
stiffness k. The knowledge of these three quantities, k, f\textsubscript{0}
and Q, is the only need for the calibration of the linear transformation.

\begin{figure}
\begin{centering}
\includegraphics[width=8.5cm]{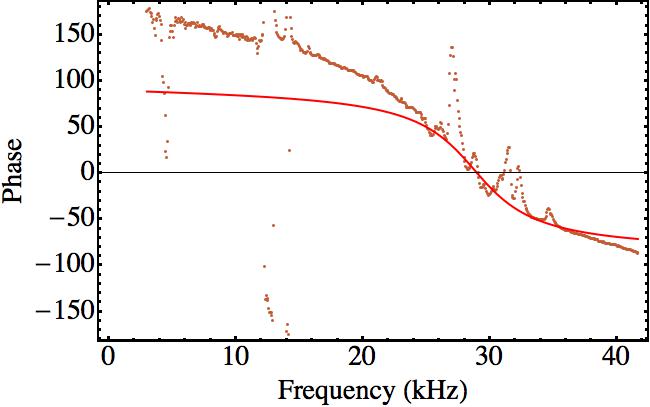}
\par\end{centering}

\protect\caption{Phase response of the cantilever in liquid far from the sample, excited
by a piezodither. The measured phase is fitted by the theoretical
phase of a harmonic oscillator to extract the resonance frequency
f\protect\textsubscript{0} and the quality factor Q.}

\end{figure}

As the tip radius is an important parameter, we glued a polystyrene
sphere of 20\textgreek{m}m of radius, coated with 40nm of gold, to
the lever. Glue gets the lever stiffer than a brand new lever. It
has stiffness of 10 N/m, as measured by Brownian motion, its first
eigenmode in salt-water is 29 kHz and its quality factor is 4. The
low quality factor gives here a large bandwidth to the system. This
gives a good signal to noise ratio even far from the eigenmode, it
is especially important for frequency above the first eigenmode. It
is important to highlight that cantilever is the same lever for all
presented experiments. If one wants to perform this experiment with
conventional dynamic AFM, one should use different cantilevers. In
order to avoid elastic-dominated behavior the sample is chosen to
be silicon with a Young modulus\cite{boyd2012measurement} of 170GPa.
We performed Z-spectroscopy in presence of silicon in salt-water solution
and recorded the amplitude n(z) and phase \textgreek{F}(z) for arbitrary
excitation frequency from 1kHz up to 40kHz. Fig.2 presents the amplitude
and phase change with respect to the tip-sample distance. The zero
is defined as the mechanical contact between the tip and the sample.
In this regime the mechanical zero is expected to match the hydrodynamic
zero. One observation is that all signals are different even if they
are due to the same interaction. As it is shown in Fig.2 the noise
in this curve is essentially due to the Brownian motion at room temperature
and to the moving liquid. It is important to notice that the signal
to noise ratio is the same whatever the excitation frequency which
is essentially due to the low quality factor.

\begin{figure}

\begin{centering}
\includegraphics[width=8.5cm]{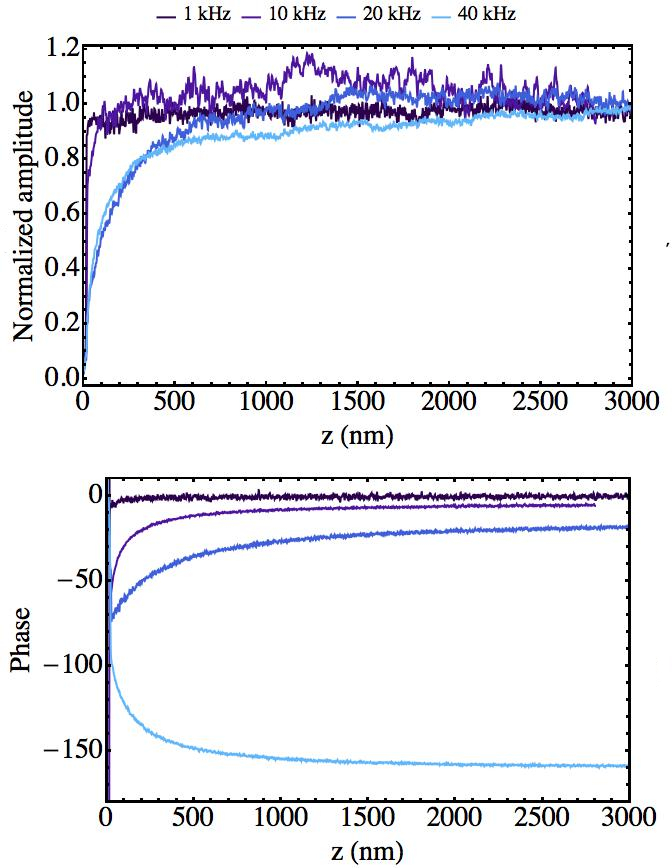}\protect\caption{ Normalized amplitudes and phase signals (as defined in text) measured
at 1kHz, 10kHz, 20kHz and 40kHz, respectively from purple to blue,
versus tip sample distance. Despite their large apparent differences,
these signals are the measurement of the same interaction (see text
and next figure). }

\par\end{centering}

\end{figure}

Fig.3 reports the measured dissipation and the stiffness obtained
through the linear transformation, using Eq.2 and 3, and the amplitude
and phase change presented in Fig.2. Measured damping is directly
compared to the theoretical damping given by lubrication approximation
(cf Eq.1) without adjustable parameters. As expected, the measured
damping matches the theoretical damping in this regime. The damping
measured at 40 kHz presents a small deviation from the theory. It
presents an non-reproducible offset between the mechanical zero and
the hydrodynamic zero of around 50nm, which might is most highly due
to the presence of an impurity. The stiffness is essentially zero
except when the tip is in mechanical contact with the sample and matches
the minus of the derivative of the static force, as shown in green
in Fig.3. This result is coherent with experiment performed in salt-water
whose ions screen the surface potential and with the lubrication theory
which gives a null stiffness for this regime following: 
\[
G'=(\frac{6\pi\omega\eta R^{2}}{z})^{2}\frac{1}{E^{*}\pi\sqrt{2RD}}
\]

Where E\textsuperscript{{*}}is the reduce Young modulus of the sample.

\begin{figure}
\begin{centering}
\includegraphics[width=8.5cm]{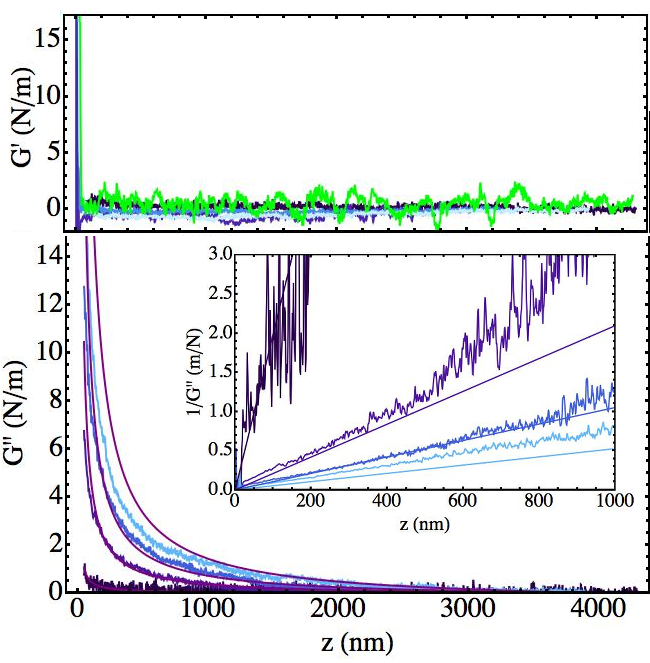}
\par\end{centering}

\protect\caption{The stiffness G' and issipation G'' , from purple to blue respectively
1kHz, 10kHz, 20kHz and 40kHz, are obtained using Eq.2 and 3 and the
amplitude and phase change presented in Fig.2. Minus of the derivative
of the static force is presented in green. Each point is a average
of 10 measurements points. The measured stiffness G' matches minus
of the derivative of the static force. The measured dissipation is
compared to the theoretical dissipation, continuous purple line, given
by the lubrication theory Eq.1. Inset, inverses of the dissipation
G'' with respect to the tip-sample distance for different excitation
frequencies are compared to the inverse of the theoretical dissipation.
The 1/G'' curves cross the distance axis at the same point, which
define the hydrodynamic surface position, which in this case matches
the mechanical zero, execpt for 40kHz as explain in the text.}

\end{figure}

To further characterize the viscous damping, we add in the inset of
Fig.3 to the inverse of the measured damping 1/G'' the inverse of
the theoretical dissipation. This demonstrates that the measured damping
behaves as d\textsuperscript{-1}. The hydrodymanic zero can be extracted
from the intersection with the z-axis. To the precision of our measurment
in the case of hard sample, the hydrodymanic zero matches the mechanical
zero. This result first shows that when in linear regime there is
no direct benefit to work at the first eigenmode especially in liquid.
Second, when linear regime is established it is easy to recover the
dynamic interaction whatever the excitation frequency around a eigenmode.
Finally, that proves that even a simple interaction can varied from
order of magnitude with respect to the frequency. The FFM is thus
a helpful tool to time dependent interaction.

In conclusion we have demonstrated that we are able to recover quantitatively
the dynamic part of the interaction in a liquid media with only the
knowledge of the transfer function of the lever in the experimental
conditions and without adjustable parameters. The good agreement of
the calculated dissipation with the experimental data shows that the
FFM is an excellent candidate for performing experiment into liquid
and open new route toward the measurement of mechanical properties
of soft matter. From the dynamic transition from a viscous-dominated
behavior at large distance to an elastic-dominated behavior at short
distance, we aim to extract the Young modulus of soft and thin film\cite{leroy2011hydrodynamic}.
The increase in the frequency range of this mechanical testing, compared
to the SFA, allows us to probe the possibility of the frequency dependence
and the viscoelastic properties of soft and thin films. 

\bibliographystyle{unsrt}
\bibliography{library}

\end{document}